\documentclass[galaxies,article,accept,oneauthor,dvi2pdf,10pt,a4paper]{Definitions/mdpi}

\usepackage{graphics,epsf}
\usepackage{amsmath}                
\usepackage{amsfonts}               
\usepackage{amssymb}                
\usepackage{epsfig}                 
\usepackage{graphicx}
\usepackage{soul,upgreek}
\usepackage{booktabs}
\usepackage{multirow}
\usepackage{microtype}

\usepackage{environ}
\NewEnviron{equation1}{%
\begin{equation}
\scalebox{0.95}{$\BODY$}
\end{equation}
}

\usepackage{subfigure}
\makeatletter
\renewcommand{\@thesubfigure}{\normalsize(\textbf{\alph{subfigure}})}
\makeatother

\def \Gyr{~\rm{Gyr}}

\firstpage{1} 
\pubvolume{6}
\issuenum{3}
\articlenumber{89}
\pubyear{2018}
\copyrightyear{2018}
\history{Received: 15 July 2018; Accepted: 9 August 2018; Published: 15 August 2018}

\Title{Jsolated Stars of Low Metallicity}

\Author{Efrat Sabach \orcidA{}}

\AuthorNames{Efrat Sabach}

\address[1]{%
Department of Physics, Technion---Israel Institute of Technology, Haifa 32000, Israel; efrats@physics.technion.ac.il\\}

\abstract{We study the effects of a reduced mass-loss rate on the evolution of low metallicity Jsolated stars, following our earlier classification for angular momentum (J) isolated stars.~By using the stellar evolution code  \texttt{MESA} we study the evolution with
different mass-loss rate efficiencies for stars with low metallicities of $Z=0.001$ and  $Z=0.004$, and compare with the evolution with solar metallicity, $Z=0.02$.~We further study the possibility for late asymptomatic giant branch (AGB)---planet interaction and its possible effects on 
the properties of the planetary nebula (PN).
We find for all metallicities that only with a reduced mass-loss rate an interaction with a low mass companion might take place during the AGB phase of the star. The interaction will most likely shape an elliptical PN.
The maximum post-AGB luminosities obtained, both for solar metallicity and low metallicities, reach high values corresponding to the enigmatic finding of the PN luminosity function.}


\begin{document}


\section{Introduction}
{\it Jsolated stars} are stars that do not gain much angular momentum along their post main sequence evolution from a companion, either stellar or substellar, thus resulting with a lower mass-loss rate compared to non-Jsolated stars \cite{SabachSoker2018b}.
As previously stated in Sabach and Soker~\cite{SabachSoker2018b,SabachSoker2018a}, the fitting formulae of the mass-loss rates for red giant branch (RGB) and asymptotic giant branch (AGB) single stars are set empirically by contaminated samples of stars that are classified as ``single stars'' but underwent an interaction with a companion early on, increasing the mass-loss rate to the observed rates.
The~mass-loss rate on the giant branches has extensive effects on stellar evolution and on the resulting planetary nebula (PN) in low and intermediate mass stars.
The reduced mass-loss rate of Jsolated stars results in a larger AGB radii compared to the RGB and compared to the ``traditional'' evolution with the high mass-loss rate efficiency of non-Jsolated stars.
The higher AGB radii reached for Jsolated stars can lead to possible late interaction with a low mass companion.
If such a Jsolated star interacts late in its evolution with a companion, thus no longer qualifying as a Jsolated star afterwards, strong interaction might cause angular momentum gain, spin up, and increase in the mass-loss rate.

The role of low mass companions (brown dwarfs or planets) in shaping PNe has been long discussed over the past few decades and it has been suggested that most PNe result from binary interaction 
(e.g., \cite{Soker1996, SiessLivio1999a, SiessLivio1999b, DeMarcoMoe2005, SokerSubag2005, MoeDeMarco2006,  VillaverLivio2007, VillaverLivio2009, Nordhausetal2010, DeMarcoSoker2011, Mustilletal2014, Villaveretal2014, Meynetetal2017}).
As we have shown in \citet{SabachSoker2018b} for solar type Jsolated stars (both in mass and in metallicity),
such an interaction can occur during the AGB phase  of evolution, where the companion is likely to be engulfed by the star.
The engulfed companion will deposit angular momentum to the primary's envelope, increasing the mass-loss rate and by that later accelerating the post-AGB evolution.
This late interaction can shape an elliptical bright PN. 
In addition, we further found under the Jsolated framework  that as the sun is a Jsolated star it will most likely engulf the earth during the AGB rather than during the RGB.

We have also shown that such Jsolated stars have implications related to the puzzle of
the bright end cut-off in the PN luminosity function (PNLF) of old stellar populations (\cite{SabachSoker2018b,SabachSoker2018a};
 for studies on the PNLF see, e.g., \cite{Ciardulloetal1989, Jacoby1989, Ciardulloetal2005, vandeSteeneetal2006, Ciardullo2010, Davisetal2018, Gesickietal2018}).
It was observed that both young and old populations have a steep bright end cut-off in the PNLF in [OIII] emission lines at $M^*_{5007}\simeq-4.5$~mag. This implies a more massive central star than expected in old populations, that reach post-AGB luminosities $L\geq 5000~L_\odot$ in order to ionize the observed bright nebulae to the desired level.
Our previous results indicate that also for low mass stars the post-AGB luminosities of Jsolated stars are bright enough to account for
the bright end cut-off in the PNLF of old stellar populations.

In \citet{SabachSoker2018a} we focused on the implications of a reduced mass-loss rate on stellar evolution of solar-type stars and the shaping of elliptical PNe by a companion.
In  \citet{SabachSoker2018b} we set the term {\it Jsolated stars} and studied the possible solution for the puzzling finding of bright PNe in old stellar populations, where the stellar mass is up to $1.2~M_\odot$. Yet, we have only focused on Jsolated stars of solar metallicity, $Z=0.02$.
Here we continue the research and study the evolution of Jsolated star with low metallicities.
\section{Results}

We continue the study of Jsolated stars in old stellar populations by studying the evolution of low and intermediate mass stars with reduced mass-loss rates and with low metallicities of $Z=0.001$ and $Z=0.004$, compatible with the old population of the Small Magellanic Cloud 
(where the metallicity has been measured to be between $Z=0.001$ and $Z=0.004$; \cite{Diagoetal2008}).
We conduct our simulations using the Modules for Experiments in Stellar Astrophysics (\texttt{MESA}, version 10398 \cite{Paxtonetal2011}).
We compare the evolution of stars with low metallicity to the evolution of stars with solar metallicity, $Z=0.02$.
\citet{Badenesetal2015}~find that most PN progenitors in the Large Magellanic Cloud correspond to stars of initial mass between $1~M_\odot$ and $1.2~M_\odot$.
We here consider stars with an initial mass as low as $0.9M_\odot$ for our study of Jsolated stars and the implications to the resulting PNe.

We focus on the effects of a reduced mass loss rate efficiency, as expressed by the empirical mass-loss formula for red giant stars of \citet{Reimers1975}
\begin{equation}
\dot{M}=\eta\times4\times10^{-13}LM^{-1}R.
\label{eq:Reimers}
\end{equation}

We repeat the procedure described in \citet{SabachSoker2018b} and follow the evolution with several mass-loss rate efficiency parameters, from $\eta=0.5$ (the ''traditional'' commonly used mass-loss rate efficiency; e.g.,  \cite{Guoetal2017}) and as low as $\eta=0.05$.

To examine whether an interaction can take place between a giant star and its low mass companion we focus on the condition for tidal capture (spiral-in of the low mass companion).
\citet{Soker2004} expressed the maximum orbital separation at which tidal interaction is significant for brown dwarfs and planets.
The maximum radius, scaled for a $10~M_J$ planet companion, is given by
\begin{equation1}
a_{\rm m}= 3.9 R\left( \frac{\tau_{\rm ev}}{6~\times~10^5{\rm yr}}\right)^{1/8}\left( \frac{L}{2000L_\odot}\right)^{1/24}
\left( \frac{R}{200R_\odot}\right)^{-1/12}
\left( \frac{M_{\rm env}}{0.5M_1}\right)^{1/8}
\left( \frac{M_{\rm env}}{0.5M_\odot}\right)^{-1/24}
\left( \frac{M_2}{0.01M_1}\right)^{1/8},
\label{eq:acapture}
\end{equation1}
where $L$, $R$ and $M$ are the luminosity, radius, and mass of the giant (RGB or AGB star), respectively, $M_{\rm env}$ is the giant's envelope mass, and $\tau_{\rm ev}$ is the evolution time on the upper RGB or AGB.
In other words, for a $10~M_J$ companion to be engulfed by the star the
condition on the ratio between the giant's radius and the orbital separation is $R/a>0.26$.

\begin{figure}[H]
\centering
\includegraphics[trim= 2.9cm 2cm 2cm 2cm,clip=true,width=0.59\textwidth]
{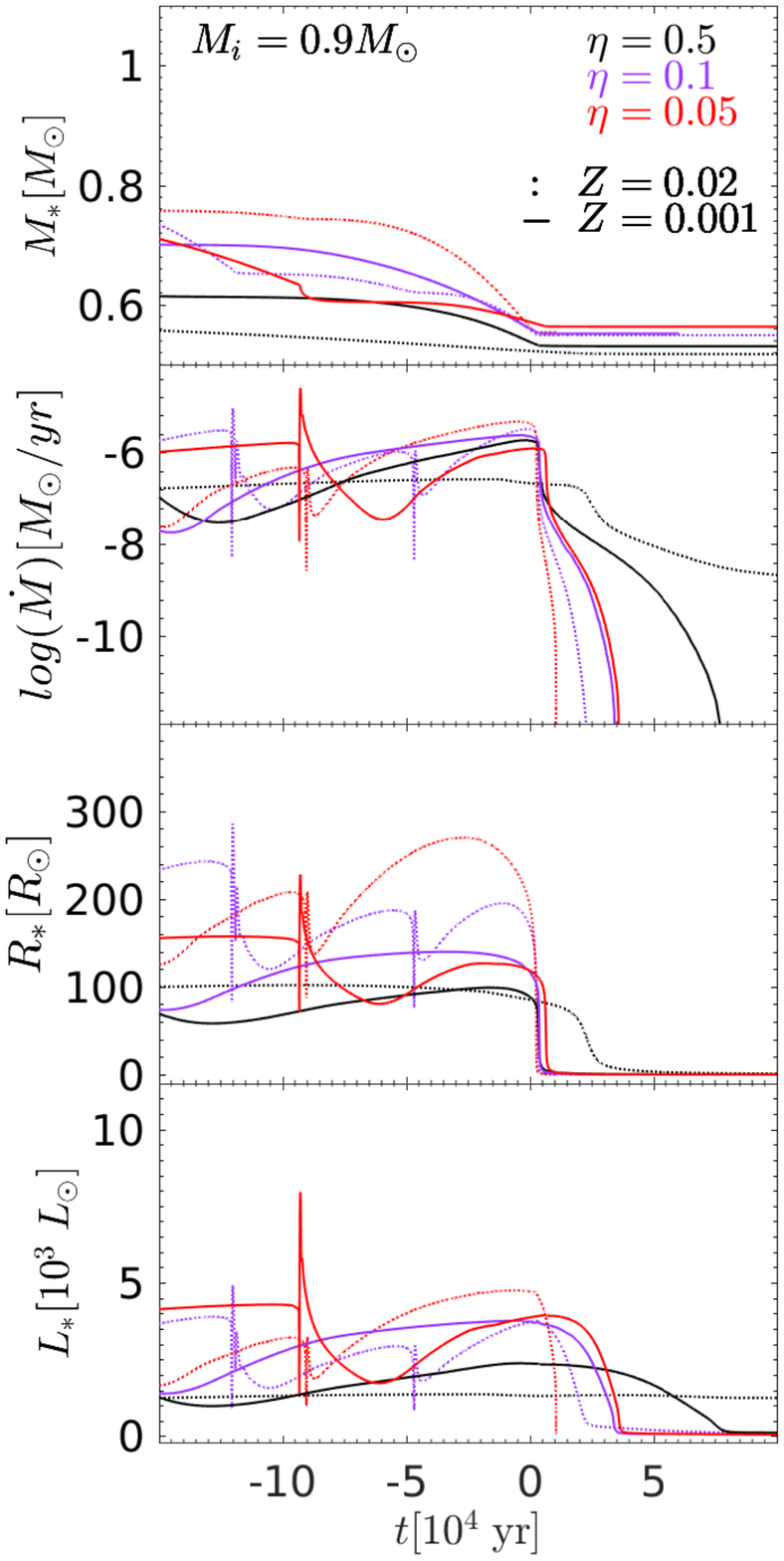}
    \hspace{-2.85cm}
\includegraphics[trim= 5.45cm 2cm 2cm 2cm,clip=true,width=0.5\textwidth]
{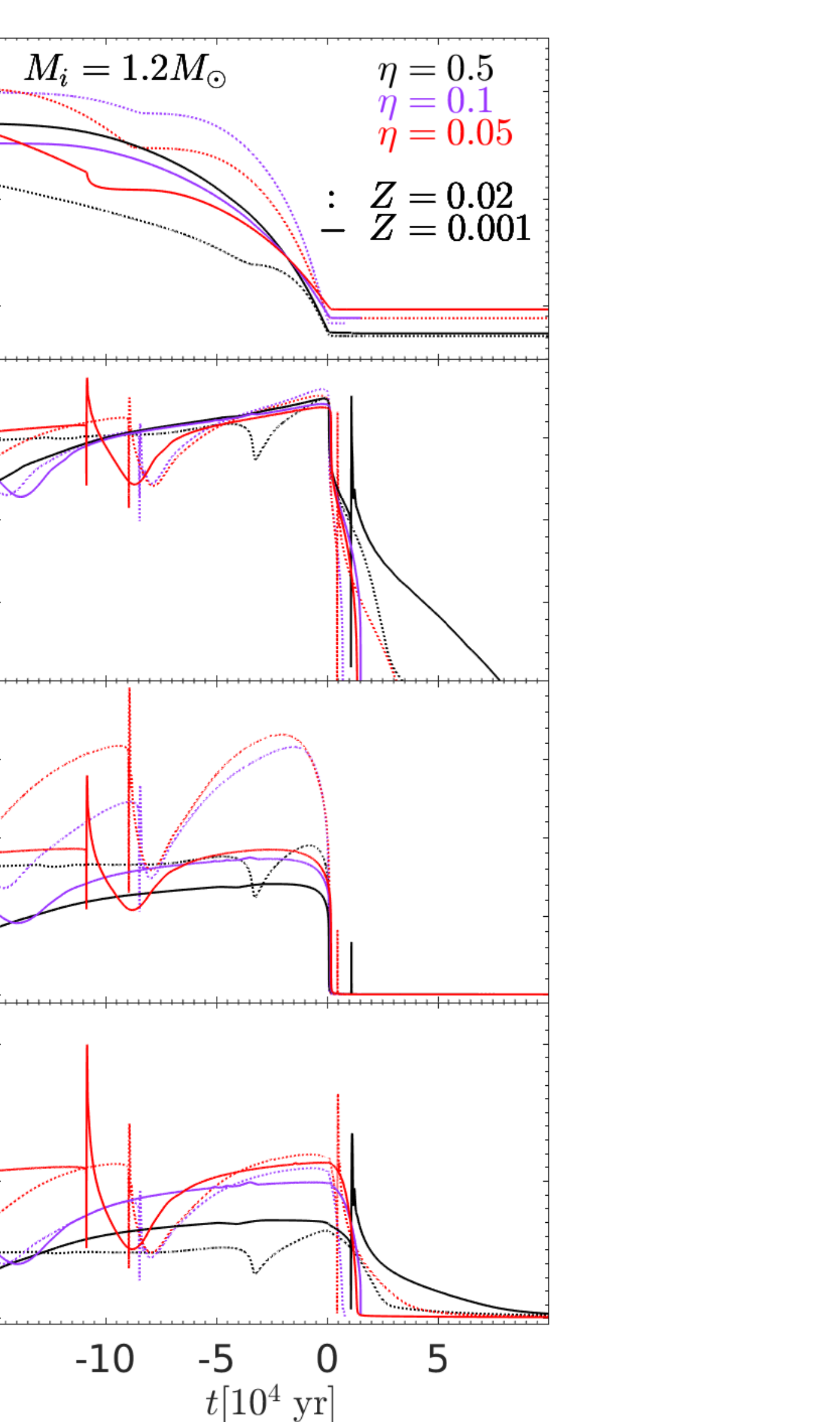}
\caption{The evolution during the final $\simeq2.5\times10^5{\rm~yr}$ of the  asymptotic giant branch (AGB) of stars of initial mass $0.9~M_\odot$ ({\bf left plot}) and $1.2~M_\odot$ ({\bf right plot}).
The graphs are shifted so that at $t=0$ the envelope mass of the star is equal to $0.01~M_\odot$.
We present the evolution for 2 	metallicities: $Z=0.02$ (solar; dotted) and $Z=0.001$ (old population; solid), and for three mass-loss rate efficiency parameters, $\eta=0.5$ (the ``traditional'' commonly mass-loss rate efficiency; black), $\eta=0.1$ (purple), and $\eta=0.05$ (red).
The panels depict, from top to bottom: the mass of the star,
the mass loss rate  (logarithmic of the absolute value), the stellar radius, and the stellar luminosity.
} 
\label{fig:pAGB}
\end{figure}

\begin{figure}[H]
\centering
\includegraphics[trim= 2.9cm 2cm 2cm 2cm,clip=true,width=0.59\textwidth]
{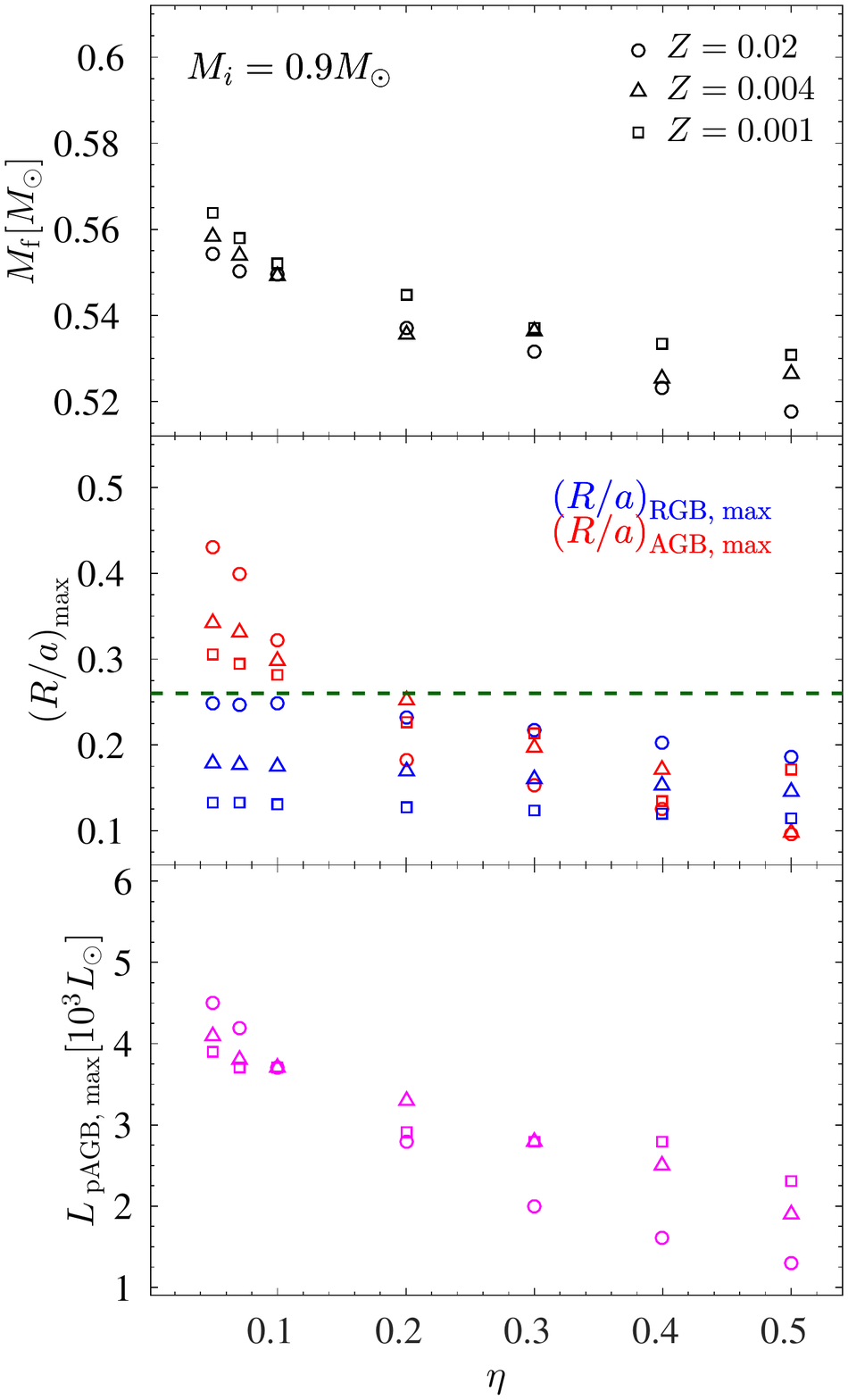}
    \hspace{-1.6cm}
\includegraphics[trim= 5.45cm 2cm 2cm 2cm,clip=true,width=0.5\textwidth]
{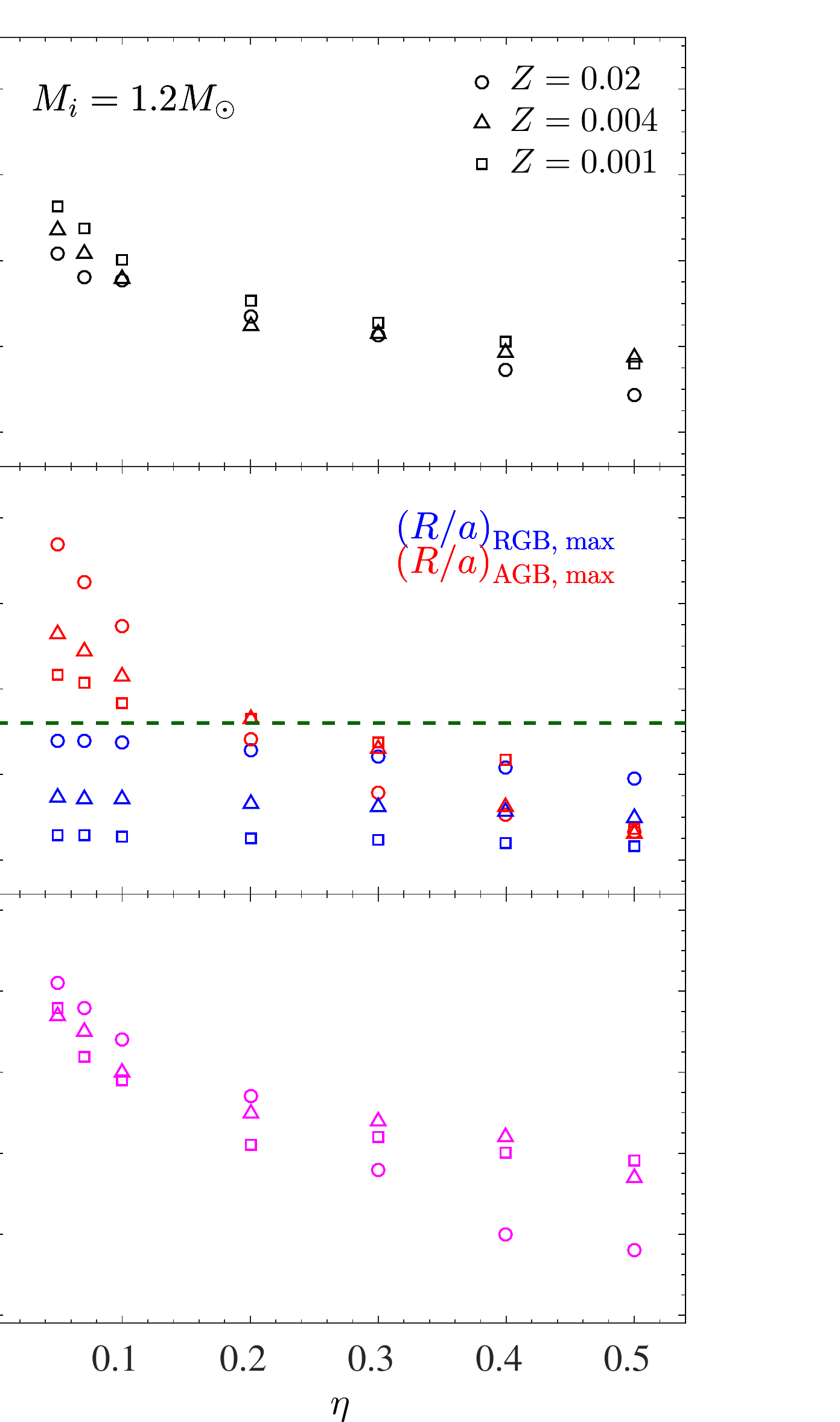}

\caption{
The summary of the evolution of a $0.9~M_\odot$ star ({\bf left plot}) and a $1.2~M_\odot$ star  ({\bf right plot}).
For~each star we studied the evolution from zero age main sequence until the formation of a white dwarf, for several mass-loss rate efficiency parameters, from $\eta=0.5$ (the ``traditional'' commonly mass-loss rate) to as low as $\eta=0.05$.
We present the evolution for 3 metallicities: $Z=0.02$ (circles), $Z=0.004$ (triangles) and Z = 0.001 (squares).
The upper panels present the final white dwarf mass of each star.
The middle panels present the maximum value of the stellar radius over the orbital separation, $\left(R/a\right)_{\rm max}$, for both the red giant branch (RGB; blue) and the asymptotic giant branch (AGB; red). The~companion mass is $10~M_J$ and the initial orbital separation taken is $3~AU$.
The~green horizontal line indicates the capture condition above which planet engulfment can take place (Equation~(\ref{eq:acapture})).
The~lower panels present the maximum value of the luminosity on the post-AGB, $L_{\rm pAGB, \;max}$.}
\label{fig:summary}
\end{figure}

In Figures \ref{fig:pAGB} and \ref{fig:summary} we present the results of our simulations for a $0.9~M_\odot$ star and a $1.2~M_\odot$ star, since these masses 
bound the relevant mass range of PN progenitors of old stellar populations. 
In~Figure  \ref{fig:pAGB} we present the final $\simeq2.5\times10^5~{\rm yr}$ of the AGB and focus on 2 metallicities, $Z=0.02$ and $Z=0.001$, and three mass-loss rate efficiency parameters, $\eta=0.5$ (the ``traditional'' commonly used mass-loss rate efficiency parameter), $\eta=0.1$, and $\eta=0.05$.
When examining the effects of metallicity and reduced mass-loss on the evolution we focus on the values at the last AGB pulse.
The plots are shifted so that they are centered around $t=0$ where $M_{\rm env}=10^{-2}~M_\odot$.
We present (top to bottom) the mass, the mass loss rate, the radius and the luminosity.
Since tidal interaction brings the planet into the envelope, and this interaction is highly sensitive to the ratio of $R(t)/a(t)$, We examine this quantity in details.
We~define the value of $L_{\rm pAGB, \;max}$ by the maximum value of the luminosity during the last AGB pulse (around $t=0$), disregarding the rapid rise in luminosity due to the helium flash.
The reasoning behind choosing this value is that the short duration of the helium flash will have a small effect on observations whereas the ``plateau'' in the final pulse better reflects the observed luminosities.
In~Figure~\ref{fig:summary} we present the important values which have implication on the shaping and on the brightness of the resulting PN (if at all observed): the final mass (upper panel), the maximum value of $R/a$ as expressed in Equation~(\ref{eq:acapture}), both on the RGB and on the AGB, and the maximum value of the post-AGB luminosity.

\section{Discussion}

We studied the evolution of Jsolated stars with low metallicities ($Z=0.001$ and $Z=0.004$) and with a reduced mass-loss rate efficiency, from $\eta=0.5$ (the commonly used value) to as low as $\eta=0.05$. We compared the evolution of Jsolated stars with low metallicity with that of Jsolated stars with solar metallicity, $Z=0.02$.
Our conclusions are as follows.
\begin{enumerate}[leftmargin=*,labelsep=4.9mm]
\item	
A higher final mass (higher mass of the central star), $M_f$, implies a more luminous central star for the ionization of the PN. It can be seen for all metallicities that as the mass-loss rate efficiency parameter $\eta$ decreases the value of $M_f$ increases.
There are very small differences between the values of $M_f$ for different metallicities (yet with the same initial mass and the same value of $\eta$).
Overall, by comparing  $M_{f, \eta=0.5}$ and $M_{f, \eta=0.05}$ there is an increase of 5--8\%.
On the one hand this increase in $M_f$ might be too small for dynamical effects, but on the other hand it has a large effect on the luminosity, as we shall discuss below, since the luminosity is very sensitive to the central star mass.

\item	The ratio of the maximum stellar radius and the orbital separation, $\left( R/a\right)_{\rm max}$, clearly increases both with the decrease in $\eta$ and with increasing $Z$. 
Moreover, when examining Equation~(\ref{eq:acapture}) for a planet companion of $10~M_J$ and at an initial orbital separation of $3~AU$ we reach the final conclusions as in \citet{SabachSoker2018a}: For the RGB phase it is only marginal for the planet to be engulfed in all cases, hence the probability for an early interaction is at most very small.
In~addition, for the traditional evolution with a high mass-loss rate efficiency parameter of $\eta=0.5$ the AGB value of $\left( R/a\right)_{\rm AGB,\; max}$ is too small for planet engulfment for the low mass of $0.9~M_\odot$, and is marginal for the larger mass of $1.2~M_\odot$.   
However, when reducing the mass-loss rate efficiency to $\eta\le 0.1$ it is clear from our results that 
there is a non-negligible range of separations for a planet to exist and be engulfed by the star during the AGB phase.
This is because the AGB radius reaches higher values compared to the RGB and compared to the values obtained in ``traditional'' evolution.
In other words, planet engulfment will take place on the AGB of Jsolated stars when the mass-loss rate efficiency parameter is $\eta \le 0.1$, independent on the metallicity, for the representative cases studied here.
\item Though we found that both for solar metallicity and for low metallicities planet engulfment is likely to take place on the AGB of Jsolated stars, it is interesting to note that the value of $\left( R/a\right)_{\rm max}$ not only increases with an increasing value of $Z$, but the increase is also ``stronger'' as $\eta$ reduces and the metallicity increases.
\item	
Our results have implications on the bright end cut-off of the PNLF in old stellar populations, where luminosities of $\simeq$5000~$L_\odot$ and higher are needed for the central star to ionize the bright nebula.
To examine this possibility of such a bright central star we focus on the post-AGB luminosities reached in our simulations.
We find that 
the post-AGB luminosity also increases with the decrease of the mass-loss rate, reaching the high $\simeq$5000~$L_\odot$ luminosities needed to explain the brightest PNe in old stellar populations. Interestingly it seems that as the value of $\eta$ decreases the values for different metallicities grow closer together.
We point out that the value of $L_{\rm pAGB, \;max}$ has a wide range since it is an approximate value taken at the final AGB phase (the maximum value of the luminosity around $t=0$ in Figure \ref{fig:pAGB}), disregarding the sharp rise due to the helium shell flash.
Indeed, we have shown that under the Jsolated framework we can explain the bright end cut-off of the PNLF also for low  metallicity low mass stars.
\end{enumerate}

Overall we reached for the case of low metallicity stars of $Z=0.001$ and $Z=0.004$
the same conclusions as in our previous works \cite{SabachSoker2018a, SabachSoker2018b}.
Jsolated stars, with a mass loss rate efficiency parameter $\eta\le 0.1$ and in the mass range of $M_{\rm i}$ = 0.9--1.2~$M_\odot$,
reach higher radii on the AGB compared to the RGB and also higher AGB radii compared to non-Jsolated stars, both for solar metallicity and low metallicities.
The post-AGB luminosities of Jsolated stars also reach higher values for all metallicities studied.
For the higher masses of $1.2~M_\odot$ we find that the luminosities reached with a mass-loss rate efficiency parameter of $\eta=0.1$ are $\simeq$5000~$L_\odot$ and might account for the PNLF in old stellar populations.
For the lower mass stars of $0.9~M_\odot$ we find that planet engulfment can take place on the AGB for a low mass loss rate efficiency parameter, but only for a very low efficiency parameter of $\eta=0.05$
the luminosity exceeds $4000~L_\odot$.

Together with a late interaction with a low mass companion such Jsolated stars could account for the shaping of elliptical PNe and possibly also the bright PNe in old stellar populations \cite{SabachSoker2018a, SabachSoker2018b}.
Once such an interaction takes place the star is no longer considered a Jsolated star.
Recently, \mbox{\citet{Giles2018}} found the longest period transiting planet candidate from radial velocity measurements, EPIC248847494b. The relevant system parameters are a planet of mass $1-10~M_J$ at $4.5\pm1.0~AU$ from a $0.9\pm0.09~M_\odot$ star. It will be interesting to examine the future of the system within our Jsolated framework.

We point out that new stellar evolution simulations find higher post-AGB luminosities compared to old calculations (e.g., \cite{Karakas2014, Bertolami2016, Venturaetal2018}).
\citet{Gesickietal2018} present new evolutionary tracks of low-mass stars and study the bright end cut-off of the PNLF.
They use a different stellar evolution code and find for populations up to an age of  
$7 \Gyr$ that the PNLF peak can be obtained by lower-mass stars than previously thought.
The full answer to the puzzle of the PNLF might be a combination of our results of a low mass-loss rate of Jsolated stars and such new 
stellar evolution calculations.





\vspace{6pt}

\funding{This research was funded from the Israel Science Foundation and a grant from the Asher Space Research Institute at the Technion.}

\acknowledgments{I thank Noam Soker  and the referees for useful comments
that helped improve the paper.}

\conflictsofinterest{The author declares no conflicts of interest.}

\reftitle{References}




\end{document}